\documentclass[aps,prx,a4paper,twocolumn,longbibliography,floatfix,grouppedaddress]{revtex4-2}
\usepackage{amssymb,amsmath,amsfonts}
\usepackage{graphicx}
\usepackage[dvipsnames]{xcolor}
\usepackage[bookmarksopen,colorlinks,linkcolor=blue,citecolor=olive,urlcolor=blue]{hyperref} 
\usepackage{bm}
\def\div{\mathop{\rm div}}

\DeclareMathOperator{\sech}{sech}

\def\tr{\mathop{\rm tr}}
\def\bj{\mathbf{j}}

\def\br{\mathbf{r}}
\def\vp{\varphi}
\def\eps{\varepsilon}
\def\D{\Delta}
\def\t{\theta}
\def\x{\chi}
\def\l{\lambda}
\def\G{\Gamma}
\def\k{\kappa}

\def\dd{\delta}

\def\sinn{\mathfrak{s}_n}
\def\cosn{\mathfrak{c}_n}

\def\be{\begin{equation}}
	\def\ee{\end{equation}}

\begin{document}
	
	\title{Thermal phase slips in superconducting films near the critical current\\
		at arbitrary temperatures}
	
	\author{Ivan M.\ Artemov}
	\affiliation{L.~D.\ Landau Institute for Theoretical Physics,  Chernogolovka 142432, Russia}
	\affiliation{Moscow Institute of Physics and Technology, Dolgoprudny 141701, Russia}
	
	\author{Mikhail A.\ Skvortsov}
	\affiliation{L.~D.\ Landau Institute for Theoretical Physics,  Chernogolovka 142432, Russia}
	\affiliation{Moscow Institute of Physics and Technology, Dolgoprudny 141701, Russia}
	\date{\today}
	
	\begin{abstract}
We develop a theory of thermal phase slips in disordered superconducting films biased near the critical current $I_c(T)$. Generalizing recent results obtained close to $T_c$, we show that the optimal fluctuation governing the phase-slip barrier in two dimensions satisfies the exactly integrable Boussinesq equation for arbitrary temperatures $T<T_c$.
Since both the transverse and longitudinal sizes of the optimal nucleus diverge as $I\to I_c(T)$, the Usadel equation for quasiparticles in the presence of a slowly varying order parameter can be solved perturbatively using a gradient expansion.
The resulting field theory for a complex order parameter is further reduced to the Boussinesq free energy for a single real field, with the coefficients expressed as Matsubara sums over the solutions of the uniform Usadel equation at $I_c(T)$. The activation barrier near $I_c$ has the asymptotic form $\Delta F(T,I\to I_c) = E(T) (1-I/I_c)^\alpha$. 
We calculate $E(T)$ over the full temperature range for both two-dimensional films ($\alpha=3/4$) and one-dimensional wires ($\alpha=5/4$).
For films, the theory is valid within a narrow 10\% window below $I_c(T)$, where the saddle-point configuration remains vortex-free. For wires, $\Delta F(T,I\to I_c)$ provides a good approximation to the activation barrier for all temperatures and currents.
\end{abstract}
	
	\maketitle

	\section{Introduction}
	\label{S:Intro}
	
A thermal phase slip is the thermally activated decay of the dissipationless supercurrent state, accompanied by the emergence of a finite voltage \cite{Little1967}. In superconducting nanowire single-photon detectors (SNSPDs) \cite{Semenov2001, Natarajan}, this process manifests as \emph{dark counts} (voltage pulses without a photon), thereby enabling direct determination of the phase slip rate \cite{Bartolf2010,Semenov2020}.

A metastable uniform supercurrent state is a local minimum of the free energy. To destroy superconductivity, the order parameter must undergo a fluctuation large enough to escape the basin of attraction, thereby generating a finite voltage and a normal current. The rate of thermal phase slips, to exponential accuracy, is governed by the free‑energy barrier $\Delta F$ that protects the uniform supercurrent state.  This barrier is determined by the static optimal fluctuation of the order parameter, which corresponds to a saddle point of the free energy.

The structure of the saddle point responsible for a phase slip hinges on dimensionality. Although additional degrees of freedom generally provide more options for lowering the energetic barrier, they also render its precise determination more challenging, owing to the complex interplay among different saddle-point types. 
In the simplest zero-dimensional (0D) realization of a current-biased Josephson junction, the activation barrier is given by the energy difference between adjacent extrema of the phase-dependent effective potential $U(\vp) = -E_J [\cos\vp + (I/I_c) \vp]$. The barrier scales as $\Delta F_\text{0D}\propto (1-I/I_c)^{3/2}$ when the current $I$ approaches the critical current $I_c$.

The theory of thermal phase slips in a one-dimensional (1D) wire near $T_c$ was formulated by Langer and Ambegaokar (LA) \cite{LA} and later refined by McCumber and Halperin \cite{McCumber}.
They identified the optimal fluctuation as a spatially localized suppression of the order-parameter magnitude $|\Delta(x)|$, extending over a length scale $L_x \sim \xi(1 - I/I_c)^{-1/4}$, where $\xi$ is the Ginzburg-Landau (GL) coherence length.
The divergence of $L_x$ at the critical current alters the 0D barrier exponent, yielding $\Delta F_\text{1D}\propto \Delta F_\text{0D} L_x \propto (1-I/I_c)^{5/4}$. 
Notably, while the LA solution yields the phase-slip activation barrier, its order parameter remains finite everywhere (except at zero current, where it vanishes at a single point). A genuine phase-slip event involving a vanishing order parameter takes place later, during the time-dependent evolution, once the system has surmounted the barrier.

The identification of the phase-slip nucleus poses a significantly greater challenge in the two-dimensional (2D) geometry relevant to experiments. 
At low currents ($I\ll I_c$), the saddle-point configuration in an infinite film is topologically nontrivial, consisting of a vortex-antivortex (VA) pair \cite{Bulaevskii} whose separation greatly exceeds the coherence length $\xi$. Upon increasing the current, this distance decreases and the two cores begin to overlap at $I\sim I_c$.
Numerical simulation for a bounded strip geometry performed in the GL region \cite{Vodolazov} demonstrates that the VA saddle-point solution does not persist up to the critical current, but instead ceases to exist at approximately $0.9\, I_c$. At higher currents, the optimal fluctuation is topologically trivial, when $|\Delta(x,y)|$ is suppressed in the center of the optimal nucleus but remains finite everywhere \cite{Vodolazov}.

The phase-slip activation energy in an infinite 2D film has recently been found analytically in the GL regime near $I_c$ \cite{SkvPol25}. Treating $1-I/I_c$ as a small parameter allows mapping the GL equations for the complex order parameter to the elliptic \emph{Boussinesq} equation for a real field, which is exactly integrable \cite{zakharov, bogdanov}. Its nontrivial finite-energy solution was then derived via Hirota's method \cite{hirota1,hirota2}. The optimal fluctuation is strongly anisotropic, exhibiting two distinct length scales: $L_x$ parallel to the current and $L_y$ perpendicular to it, given by
\be
\label{LxLy}
  L_x\sim\xi(1-I/I_c)^{-1/4}, 
\qquad
  L_y\sim\xi(1-I/I_c)^{-1/2} .
\ee
While the longitudinal scale $L_x$ matches the LA result, the transverse size is parametrically larger: $L_y \gg L_x$. The activation barrier scales with the product of these two lengths, yielding $\Delta F_\text{2D} \propto \Delta F_\text{0D} L_x L_y \propto (1-I/I_c)^{3/4}$.

Given that SNSPDs typically operate well below $T_c$ \cite{Semenov2001, Natarajan, kitaygorsky}, generalizing the phase-slip rates obtained within the GL model to arbitrary temperatures is of practical importance. Theoretically, however, this is a demanding endeavor: the simple GL equations must be replaced by the significantly more intricate Usadel equations for the quasiclassical Green's functions at Matsubara energies, supplemented by the self-consistency condition for the order parameter. For arbitrary currents, even a numerical search for the saddle point in this coupled system proves formidable.

In this work, we demonstrate that, near $I_c$, the optimal phase-slip nucleus in an infinite superconducting film obeys the Boussinesq equation \emph{for all temperatures} $T<T_c$. Exploiting the divergence of both $L_x$ and $L_y$ as the critical current is approached, we perform a gradient expansion of the Usadel equations for a slowly varying order parameter. This procedure enables us to integrate out the quasiparticle degrees of freedom [the field $Q(\br)$], yielding a free-energy functional that depends solely on the order parameter. 
Using the near-critical condition again, this model is further reduced to a Boussinesq field theory for a single real-valued function $f(x,y)$ governed by the energy functional
\be
\label{Bouss1}
  F_\text{B}
  = C \int dx \, dy 
  \left( 
    c_1 f_{xx}^2 + c_2 f_{xy}^2 + c_3 f_{xxx}^2 + c_4 f_{xx}^3
  \right) .
\ee
The Boussinesq free energy \eqref{Bouss1} is characterized by four temperature-dependent coefficients $c_i(T)$, given by Matsubara sums of the spectral angles found by solving the uniform-current Usadel equations. For arbitrary temperatures, these coefficients are evaluated numerically. 
In this way, we obtain the asymptotic behavior of the phase-slip barrier $\Delta F$ in the limit $I\to I_c$ for all temperatures $T<T_c$.

The paper is organized as follows. In Sec.\ \ref{S:Model} we introduce the model and describe the uniform-current state. The gradient expansion of the Usadel equations is implemented in Sec.\ \ref{S:Expansion}, where we derive the free-energy functional in the Boussinesq form at nearly critical currents. Final expressions for the activation barrier in the 1D and 2D cases are presented in Sec.\ \ref{S:Activation energy}. Finally, Section \ref{S:Summary} summarizes the key ideas and main results of our analytic approach. The technical details are relegated to Appendixes \ref{S:Appendix A} and \ref{S:Appendix B}.

\section{Model}
\label{S:Model}

\subsection{Free energy of a dirty superconductor}

Theoretical description of the thermodynamic properties of a dirty superconductor is based on the replica sigma model in the Matsubara representation \cite{finkelstein, belitz}. The model is formulated in terms of the electron matrix field $Q(\br)$ and the complex order parameter field $\Delta(\br)=|\Delta(\br)|e^{i\vp(\br)}$. Since the saddle point responsible for the phase-slip activation barrier is static, we may take $Q(\br)$ to be diagonal in Matsubara energy space and $\Delta(\br)$ to be independent of imaginary time. As we are not interested in weak localization effects, we take the replica number to be one. Then the free energy acquires the form
	\begin{multline}\label{freeenergy}
		F
		=
		\frac{\pi\nu}{4}
		T \sum_n
		\int d\br \,
		\tr \left[
		D (\nabla Q_n)^2 - 4 (\epsilon_n\tau_3 + \hat\Delta) Q_n
		\right]
		\\ 
		{}+
		\frac{\nu}{\lambda} \int d\br \, |\Delta|^2,
	\end{multline}
where $\epsilon_n=2\pi T(n+1/2)$ is the Matsubara energy (the sum is limited to $|\epsilon_n|<\omega_D$), the quasiclassical Green function $Q_n(\br)\in S^2$ can be written in terms of the spectral angle $\theta_n(\br)$ and the phase of the anomalous component $\x_n(\br)$ as
\be
	Q_n
	=
	\begin{pmatrix} 
		\cos\theta_n & e^{i\chi_n}\sin\theta_n \\ 
		e^{-i\chi_n}\sin\theta_n & - \cos\theta_n
	\end{pmatrix}, 
\ee
and
\be
	\hat\Delta 
	=
	\begin{pmatrix} 
		0 & \Delta \\ \Delta^* & 0
	\end{pmatrix} .
\ee

Varying the energy functional \eqref{freeenergy} with respect to $\theta_n(\br)$ and $\x_n(\br)$ yields a pair of Usadel equations:
\begin{subequations}
\label{usadel}
	\begin{multline}\label{usadel1}
		-\frac{D}{2}\nabla^2 \t_n + \frac{D}{2}(\nabla \x_n)^2 \sin\t_n \cos\t_n 
		\\ {}
		+ \epsilon_n \sin\t_n - |\D|\cos(\x_n-\vp)\cos\t_n = 0 ,
	\end{multline}
\vspace{-7mm}
	\be\label{usadel2}
	\frac{D}{2} 
    \nabla (\sin^2\t_n \nabla \x_n) 
	=
	|\Delta|\sin(\x_n-\vp)\sin\t_n ,
	\ee
\end{subequations}
while the derivative with respect to $|\Delta(\br)|$ results in the self-consistency equation
	\be\label{selfcons}
	|\D|=\pi \l T \sum\limits_n \sin\t_n\cos(\x_n-\vp) .
	\ee
The supercurrent density is given by
	\be\label{current}
	\bj = 2e\pi \nu D T \sum_n \sin^2 \t_n \nabla \x_n .
	\ee

	Note that the variation of the free energy \eqref{freeenergy} with respect to the superconducting phase $\vp(\br)$ yields an equation that, together with Eq.\ \eqref{usadel2}, enforces current conservation, $\div \bj = 0$.

\subsection{Uniform current}
\label{SS:uniform}

A uniform supercurrent state with $\bj=j \mathbf{e}_x$ corresponds to a constant $|\Delta(\br)|=\Delta_\G$ and linearly growing phase of the order parameter $\vp(\br)=\k x$, which coincides with the quasiparticle phase: $\x_n(\br)=\vp(\br)$.
The spectral angle $\t_n(\br)=\theta_{\Gamma n}$ is space-independent and satisfies the trigonometric equation
\be\label{uniformusadel}
	\epsilon_n \sin\t_{\G n} - \D_\G \cos \t_{\G n} +\G \sin\t_{\G n}\cos\t_{\G n}=0,
	\ee
where
\be
\label{Gamma-def}
  \G=D\kappa^2/2
\ee
is the depairing rate arising from the condensate motion. 
Equation \eqref{uniformusadel} is the generic form for weak pair breaking characterized by the rate $\Gamma$ \cite{AG1961,Maki1969}.

Due to the complicated structure of Eqs.\ \eqref{selfcons}, \eqref{current} and \eqref{uniformusadel}, determining $\Delta_\Gamma$ and $\Gamma$ for a given current density $j$ and temperature $T$ requires numerical simulation \cite{Clem2012}. 
One first solves a coupled system of Eq.\ \eqref{uniformusadel} and the self-consistency equation \eqref{selfcons} with $\chi_n = \vp$, to obtain the order parameter modulus $\Delta_\G$ and the uniform spectral angles $\theta_{\G n}$ for fixed $T$ and $\G$. Then, using Eq.\ \eqref{current}, one calculates the current $j(T,\Gamma)$ and identifies the critical current density $j_c(T)$ as the maximum of $j(T,\Gamma)$ with respect to $\Gamma$ attained at some $\Gamma_c(T)$. The same procedure has recently been carried out in Ref.\ \cite{Polkin2026}.
The values of $\G_c(T)$, $\D_{\G_c}(T)$ and $\theta_{\G_c, n}(T)$ numerically obtained in this way will be used for calculating the coefficients in the free-energy expansion at the critical current [see Eq.\ \eqref{F23} and Appendix \ref{S:Appendix A}].

\section{Boussinesq free energy}
\label{S:Expansion}

\subsection{Integrating out quasiparticles near $I_c$}

The instanton configuration responsible for the phase-slip activation energy is a spatially inhomogeneous solution of Eqs.\ \eqref{usadel} and \eqref{selfcons} that asymptotically approaches, as $\br\to\infty$, the constant-$\bj$ solution described in Sec.\ \ref{SS:uniform}.
For an arbitrary current, finding the instanton is analytically intractable even near $T_c$, where the superconductor is described by a much simpler Ginzburg-Landau equation solely for $\Delta(\br)$. Nor is it possible for the coupled Usadel and self-consistency equations.

The recent progress in the analytical theory of phase slips near the critical current in the Ginzburg-Landau region \cite{SkvPol25} relies on two key observations: (i) the instanton merges with the uniform solution at $I=I_c$, and (ii) its size diverges as $I\to I_c$. The former allows to develop a perturbative expansion in the deviation from the constant-$\bj$ state, while the latter ensures the effective theory acquires a local form of the Boussinesq type.

Quite natural, both observations are not limited to $T\approx T_c$, but hold for arbitrary temperatures. This is related to a general behavior of the free energy as a function of the superfluid momentum, with a minimum corresponding to a uniform supercurrent disappearing at $I=I_c$. Therefore at $I\to I_c$ and arbitrary $T$ we may seek the instanton solution in the form
\be\label{deltaphi}
  |\Delta(\br)| = \Delta_{\G}[1+\delta(\br)], 
\qquad
  \vp(\br) = \kappa x + \phi(\br) ,
\ee
where the dimensionless functions $\delta(\br)$ and $\phi(\br)$ (i) vanish at criticality, and (ii) vary at parametrically large scales
$L_{x,y}\gg\xi(T)$ given by Eq.\ \eqref{LxLy}.

Both properties are crucial for analytical treatment of the Usadel equations \eqref{usadel} in the field of the order parameter \eqref{deltaphi}. Since the spatial response of quasiparticles is governed by the zero-temperature coherence length $\xi(0)\ll L_{x,y}$, the spectral angle $\theta_n(\br)$ and the phase of the anomalous Green function $\chi_n(\br)$, in the leading approximation, are determined by the \emph{local} value of $\Delta(\br)$. Subleading corrections can be obtained using the gradient expansion in $\nabla\delta$ and $\nabla\phi$. Furthermore, the smallness of the deviation $\Delta(\br)-\Delta_\Gamma e^{i\kappa x}$ implies that the zero-dimensional version of the nonlinear Usadel equations \eqref{usadel} (with the gradients discarded) can be solved perturbatively. The described procedure generates $\theta_n(\br)$ and $\chi_n(\br)$ as a series in $\delta(\br)$ and $\phi(\br)$ and their derivatives:
	\begin{subequations}
		\label{tx-via-df}
		\begin{multline}
			\label{t-via-df}
			\theta_n 
			-
			\theta_{\Gamma n}
			=
			\frac{\Delta_\G \cosn}{\mathcal{E}_n} 
			\delta
			-\frac{D \kappa \, \sinn \cosn}{\mathcal{E}_n}
			\phi_x
			\\{}
			+
			\frac{D\Delta_\Gamma \cosn}{2 \mathcal{E}_n^2}
			( \delta_{xx} + \delta_{yy} )
			-
			\frac{2 D \Gamma \sinn \cosn^3 }{\mathcal{E}_n^2} \delta_{xx}
			+\dots ,
		\end{multline}
		\vspace{-3mm}
		\begin{multline}
			\label{x-via-df}
			\chi_n 
			-
			\vp
			=
			\frac{D \kappa \cosn^2}{\mathcal{E}_n} \delta_x
			+
			\frac{D \sinn}{2 \Delta_\Gamma} (\phi_{xx}+\phi_{yy})
\\{}
			-
			\frac{2 D \Gamma \sinn \cosn^2 }{\mathcal{E}_n \Delta_\Gamma }
			\phi_{xx}
			+
			\dots ,
		\end{multline}
	\end{subequations}
where 
$\mathcal{E}_n=\epsilon_n \cosn + \Delta_\G \sinn + \Gamma (\cosn^2-\sinn^2)$ and we introduced the notations $\sinn=\sin\t_{\Gamma n}$ and $\cosn=\cos\t_{\Gamma n}$.
In Eqs.~\eqref{tx-via-df} we present only terms that are linear in $\delta$ and $\phi$ and contain at most two derivatives. Expressions for nonlinear terms are rather lengthy and can be obtained systematically using the standard perturbation theory.

As $I\to I_c$, only a finite number of terms in these series should be retained, governed by the scaling of the Boussinesq instanton with the deviation from criticality. According to Ref.\ \cite{SkvPol25},
	\be\label{scaling}
	\delta\sim\eps, \quad
	\phi\sim\sqrt\eps, \quad
	x/\xi\sim1/\sqrt\eps, \quad
	y/\xi\sim1/\eps .
	\ee
The scaling \eqref{scaling} implies that the expansion \eqref{t-via-df} for $\theta$ contains only integer powers of $\eps$, whereas the expansion \eqref{x-via-df} for $\chi$ contains only half-integer powers:
\begin{subequations}
\label{txexpansion}
\begin{align}
  \theta_n & = \theta_{\G n}+\theta_n^{(1)} + \theta_n^{(2)} + \dots ,
\\
  \x_n & = \vp + \chi_n^{(3/2)} + \chi_n^{(5/2)} + \dots ,
\end{align}
\end{subequations}
with $\theta_n^{(k)}, \, \chi_n^{(k)} \sim \eps^k$ (note that $\vp$ already contains the $\eps^{1/2}$ term). Substituting Eqs.\ \eqref{txexpansion} into Eq.\ \eqref{freeenergy}, we obtain the free energy as a local functional of the fields $\delta(\br)$ and $\phi(\br)$.

To find the instanton in the vicinity of the critical current, it is sufficient to determine the free energy with the accuracy $\eps^3$ \cite{SkvPol25}. In this order, we need to know only $\t_n^{(1)}$, $\t_n^{(2)}$, and $\chi_n^{(3/2)}$ (the coefficients of $\t_n^{(3)}$ and $\chi_n^{(5/2)}$ in the free energy expansion are its first variations over $\t$ and $\x$, respectively, evaluated at the uniform solution, and thus equal to zero).

As a result, we obtain the following expression for the free-energy in terms of the deviations $\delta(\br)$ and $\phi(\br)$, written in terms of the dimensionless coordinates normalized by the coherence length $\xi(T)$:
	\be
	\label{F23}
	F^{(2,3)}
	=\nu \Delta_{\G}^2 d\xi^2 \int dx \, dy \left[\sum_{i=1}^3 a_i V_i^{(2)} + \sum_{i=1}^8 b_i V_i^{(3)} \right] ,
	\ee
where $V^{(k)}$ are all allowed (respecting the symmetry $y \to -y$ and depending on $\phi$ only through $\nabla\phi$) combinations of the fields and their derivatives that scale as $\eps^k$:
\begin{subequations}\label{V23}
\begin{gather}
	V^{(2)}
	=
	(\delta^2 , \delta\phi_x , \phi_x^2 ),
\\{}
	V^{(3)}
	=
	(
	\phi_y^2 ,
	\delta \phi_x^2 ,
	\phi_x^3 ,
	\delta^3 ,
	\delta^2 \phi_x ,
	\delta_x^2 ,
	\phi_{xx}^2 ,
	\delta_{xx} \phi_x 
	) .
\end{gather}
\end{subequations}
The dimensionless coefficients $a_i$ and $b_i$ can be obtained from Eqs.\ \eqref{freeenergy} and \eqref{tx-via-df} as specific functions of $\theta_{\Gamma n}$ summed over Matsubara energies; their explicit form is presented in Appendix \ref{S:Appendix A}. 

To make the theory dimensionless, one must adopt a particular definition of the coherence length. 
Near $T_c$ it is conventionally taken as
\be
\label{xiGL}
  \xi_\text{GL}(T) = \sqrt{\frac{\pi D}{8 (T_c-T)}} ,
\ee
but its extension to arbitrary temperatures is not unique. 
One possibility is to relate $\xi(T)$ to the upper critical field via
$H_{c2} = \Phi_0/2\pi \xi^2(T)$, where $\Phi_0$ is the flux quantum.
Using the known result for $H_{c2}$~\cite{Maki1964,HW1965} yields the implicit equation $\ln T_c/T=\psi(1/2+D/4\pi T\xi^2)-\psi(1/2)$,
with $\psi$ being the digamma function. However, this definition relies on a normal‑state property rather than on the superconducting state itself.
A more intrinsic choice is to define $\xi(T)$ through the static amplitude (Higgs) fluctuation propagator
(see, e.g., Eq.\ (12) of Ref.\ \cite{LO1971} or Eq.\ (58) of Ref.\ \cite{SF2013}), either from its pole or a low-momentum expansion.

In our analysis we choose $\xi(T)$ to coincide with $\xi_\text{GL}(T)$. Being the simplest, this choice also facilitates comparison with previously available results in the GL region \cite{SkvPol25}.

\subsection{Reduction to one real field}

The free energy \eqref{F23} written in terms of (corrections to) the order parameter modulus and phase should now be solved for the inhomogeneous saddle point. 
To this end, we start with the current conservation relation $\delta F/\delta\phi=0$, which yields
\be
\label{divj=0}
  \div\bj\propto-a_2\delta_x-2a_3\phi_{xx}+\dots=0 .
\ee 
Although this equation looks rather formidable, a crucial simplification is achieved by invoking the known scaling relations \eqref{scaling}. 
They allow us to establish a relation between the coefficients of the $\eps$-expansions for $\delta(\br)$ and $\phi(\br)$, enabling both quantities to be expressed in terms of a single real field $f(\br)$.
Since this field plays a role analogous to that of a gauge field, its precise introduction is largely a matter of convenience. Accordingly, choosing a gauge in which the phase $\phi(\br)$ is represented in the form%
\begin{subequations}
\label{reduction}
\be\label{reductionphi}
  \phi=-A f_x ,
\ee
we obtain the following expression for $\delta(\br)$:
\be\label{reductiond}
  \delta = r_1 f_{xx}+r_2 f_{yy}+r_3 f_{xx}^2+r_4 f_{xxxx} + \dots
\ee
\end{subequations}
In the representation \eqref{reduction}, $f(\br) \sim O(\eps^0)$, the terms omitted in Eq.\ \eqref{reductiond} are $O(\eps^3)$, and the coefficients $r_i$ are given by
\begin{subequations}
\begin{gather}
  r_1=2Aa_3/a_2,\\
  r_2=2Ab_1/a_2,\\
  r_3=A^2(4a_{2}a_{3}b_{2}-3a_{2}^{2}b_{3}-4a_{3}^{2}b_{5})/a_2^3,\\
\label{r4}
  r_4=-2A(a_{2}b_{7}+a_{3}b_{8})/a_2^2.
\end{gather}
\end{subequations}	
Choosing $A=\kappa\xi$ as the normalizing coefficient in Eqs.\ \eqref{reduction} ensures that our function $f(\br)$ coincides with that of Ref.\ \cite{SkvPol25} with the accuracy $O(\eps)$ (as will become apparent from the expression for the current derived below). 

A difference between the present approach and that used in Ref.~\cite{SkvPol25} is the gauge choice. 
The way $f$ is introduced in Ref.\ \cite{SkvPol25} makes \eqref{reductionphi} an infinite series, whereas in our case it terminates at the first term.
Note also that, unlike in Ref.~\cite{SkvPol25}, our expression \eqref{reductiond} for $\delta$ includes an extra term $f_{xxxx}\sim O(\eps^2)$. This term is proportional to the coefficient $r_4$, which vanishes as $1-T/T_c$ [see Eqs. \eqref{r4} and \eqref{aibiGL}] and therefore is irrelevant in the Ginzburg-Landau limit.
   	
Plugging now Eqs.\ \eqref{reduction} into \eqref{F23}, we obtain the free energy in the Boussinesq form \eqref{Bouss1} expressed in terms of the field $f(\br)$.
The omitted terms in the free energy density \eqref{Bouss1} are $o(\eps^3)$, the overall energy scale $C=\nu \Delta_{\G_c}^2 d\xi^2(T)$, and the coefficients $c_i$ are expressed in terms of $a_k$ and $b_l$. Like in the GL region \cite{SkvPol25}, their dependence on $I$ is essentially different. While the coefficient 
	\begin{subequations}
		\label{c1234}
		\be
		\label{c1}
		c_1
		=
		A^2 a_3 (4 a_1 a_3 - a_2^2)/a_2^2 
		\approx \gamma_1 \sqrt{1-I/I_c}
		\sim \eps
		\ee
vanishes at criticality (see Appendix \ref{S:Appendix B}), other coefficients in Eq.\ \eqref{Bouss1} remain finite at $I_c$.  Therefore it suffices to evaluate them at $I=I_c$:
		\begin{gather}
			c_2 = A^2 b_1,
			\\
			c_3 = A^2 (4 a_1^2 b_7+2 a_2 a_1 b_8+a_2^2 b_6)/4 a_1^2,
			\\
			c_4 = A^3
			(4 a_1^2 a_2 b_2 - 8 a_1^3 b_3 + a_2^3 b_4 - 2 a_1 a_2^2 b_5)/8 a_1^3 .
		\end{gather}
	\end{subequations}

Finally, rescaling the coordinates 
\be
  x=c_3^{1/2}c_1^{-1/2} \bar{x}, 
  \quad
  y=c_2^{1/2}c_3^{1/2} c_1^{-1} \bar{y}
\ee
 and the function $f=(2c_3/3c_4) \bar f$, we rewrite the free energy \eqref{Bouss1} as
\be
\label{Bouss2}
  F_\text{B}
  =
  \frac{8 c_1^{3/2} c_2^{1/2} c_3}{9 c_4^2} C S ,
\ee
where $S$ is the canonic dimensionless form for the Boussinesq free energy \cite{SkvPol25}:
\be
\label{F-Bouss-dimless}
	S = 
	\int d\bar x\, d\bar y
	\left(
	\frac12 \bar f_{\bar x\bar x}^2 
	+ \frac12 \bar f_{\bar x\bar y}^2
	+ \frac12 \bar f_{\bar x\bar x\bar x}^2
	+ \frac13 \bar f_{\bar x\bar x}^3
	\right) .
\ee

\section{Activation energy}
\label{S:Activation energy}
\subsection{Films}

In 2D, the lowest-energy nontrivial saddle point for the Boussinesq free energy \eqref{F-Bouss-dimless} is given by
\be
\label{f-2D-res}
  \bar f = - 6 \ln(\bar x^2+\bar y^2+3) ,
\ee
with $S=8\pi$ \cite{SkvPol25}. So one readily obtains the temperature dependence of the phase slip barrier in an infinite film:
\be\label{filmbar}
  \Delta F_\text{2D}
  =
  \frac{64\pi c_1^{3/2} c_2^{1/2} c_3}{9 c_4^2} C ,
\ee
where the coefficients $c_i(T)$ are given by Eqs.\ \eqref{c1234}.

It is instructive to rewrite Eq.~\eqref{filmbar} in terms of the superconducting condensation energy density $\eps_\text{cond}(T)$. 
Given that we adopt the GL expression for $\xi(T)$, we consistently use the analogous expression for the condensation energy:
\be
\label{e-cond-GL}
  \eps_{\text{cond}}(T)=\frac{4\pi^2 \nu (T_c-T)^2}{7\zeta(3)} .
\ee
Then the phase-slip activation barrier acquires the form
\be\label{ans2D}
  \Delta F_\text{2D}
  =
  c_{2\text{D}}(T) \eps_\text{cond}(T) d \xi^2 (T) (1-I/I_c)^{3/4} ,
\ee
where the dimensionless coefficient $c_\text{2D}(T)$ is given by
\be\label{c2D-gen}
  c_{2\text{D}}(T)
  =
  \frac{112\zeta(3)}{9\pi}\frac{\gamma_1^{3/2}c_{2}^{1/2}c_{3}}{c_{4}^{2}}\frac{\Delta_{\G_{c}}^{2}(T)}{(T_{c}-T)^{2}},
\ee
and the function $\gamma_1(T)$ determines the leading behavior $c_1\approx \gamma_1 \sqrt{1-I/I_c}$ near the critical current, see Eq.\ \eqref{c1}. 

	\begin{figure}
		\centering
		\includegraphics[width=\linewidth]{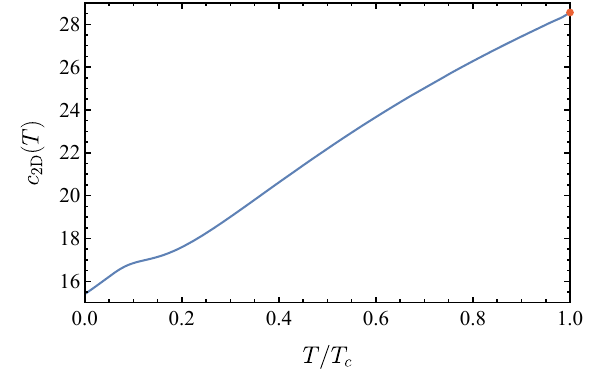}
		\caption{Temperature dependence of the coefficient $c_{2\text{D}}(T)$ in the activation barrier \eqref{ans2D} for a 2D film. The red point corresponds to the analytical result in the GL region: $c_{2\text{D}}(T_c)=2^{27/4} 3^{-9/4} \pi \approx 28.5$ \cite{SkvPol25}.}
		\label{fig:c2D}
	\end{figure}

The temperature dependence of $c_{2\text{D}}(T)$ obtained numerically is shown in Fig.\ \ref{fig:c2D}. 
At the transition, it agrees with the 
result of Ref.\ \cite{SkvPol25}: $c_\text{2D}(T_c)=2^{27/4} 3^{-9/4} \pi \approx 28.5$. At zero temperature, we find $c_\text{2D}(0) \approx 15.4$. Notably, the overall dependence is nearly linear, interpolating between these two endpoint values.

Expression \eqref{ans2D} for the activation barrier was obtained for an infinite 2D film without edges. For a strip of sufficiently large ($w \gg L_y$) but \textit{finite} width, the saddle-point configuration with the lowest activation energy is a half-instanton centered at one of its lateral edges. The thermal phase-slip barrier is then $\Delta F_{2\text{D}}/2$ \cite{SkvPol25}.

\subsubsection*{Current distribution}

It is instructive to look at an inhomogeneous current distribution at the instanton solution. Plugging the perturbative expansion \eqref{txexpansion} into the current density \eqref{current} and expanding around the uniform value $\bj=j_\G\mathbf{e}_x$ [where $j_{\G} \equiv j(T,\G)$ is defined in Sec.\ \ref{SS:uniform}], we obtain
	\be\label{curexpansion}
	\bj = j_{\Gamma} \mathbf{e}_x (1+J_1+J_2+\dots) 
	+ j_{\Gamma} \mathbf{e}_y (J_{3/2} + \dots),
	\ee
	where $J_k \sim \eps^k$ and the first three corrections read 
	\begin{subequations}\label{Ji}
		\begin{gather}
			J_1=s_1 \dd + s_2 \phi_x,
			\\
			J_{3/2}=s_3 \phi_y, \\
			J_2=s_{4}\dd\phi_{x}+s_{5}\phi_{x}^{2}+s_{6}\dd^{2}+s_{7}\phi_{xxx}+s_{8}\dd_{xx} .
		\end{gather}
	\end{subequations}
The coefficients $s_i$ are expressed in terms of the Matsubara sums $a_i$ and $b_i$ entering the free-energy expansion~\eqref{F23} as
	\be\label{scoefs}
	s=(a_{2},2a_{3},2b_{1},2b_{2},3b_{3},b_{5},-2b_{7},b_{8})/2Ab_{1}.
	\ee
	Substituting Eqs.\ \eqref{reduction} into Eqs.\ \eqref{curexpansion} and \eqref{Ji} yields the supercurrent density in terms of the field $f(\br)$:
		\be\label{j-f}
		\bj = j_\G \mathbf{e}_x (1+f_{yy}+\dots) + j_\G \mathbf{e}_y (-f_{xy} + \dots) .
		\ee

	The terms presented in Eq.~\eqref{j-f} describe the current expansion up to $O(\eps^2)$ and reproduce verbatim the corresponding expression from Ref.\ \cite{SkvPol25}. However, the local representation of the phase $\phi(\br)$ in terms of $f(\br)$ [Eq.~\eqref{reductionphi}] produces higher-order (and generally nonlocal) terms $o(\eps^2)$ in the expression for the current \eqref{j-f}, which were absent in Ref.~\cite{SkvPol25}. In turn, the local representation of the current, adopted in Ref.~\cite{SkvPol25} by introducing the field $f(\br)$ through the stream function, would lead to a nonlocal expression for the phase.
	These two constructions generate functions $f(\br)$ that coincide to the leading order, $O(\eps^0)$, but differ at higher orders, as has already been noted above.
Nevertheless, such an ambiguity, sensitive to the specific definition of $f(\br)$, does not affect the results obtained within the accuracy $O(\eps^3)$, corresponding to the Boussinesq field theory. The latter is applicable for describing only the \emph{leading} asymptotic behavior of the instanton near the critical current. This reasoning also indicates the irrelevance of the subleading, scaled as $\eps^2$, terms in the expression \eqref{reductiond} for the order-parameter correction $\delta(\br)$ (whose explicit derivation in terms of the field $f(\br)$ was merely a technical trick allowing us to obtain the leading-order corrections to the current).
	
Using Eqs.\ \eqref{c1}, \eqref{f-2D-res} and \eqref{j-f}, we obtain the instanton current distribution, which, expressed in dimensional units, is given by
	\begin{subequations}
		\begin{gather}
			\frac{j_x}{j_\Gamma} = 1-\frac{8\eps^{2} \xi^2\gamma_{1}^{2}c_{3} (\gamma_{1}c_{2}\eps x^{2}-\gamma_{1}^{2}\eps^{2}y^{2}+3c_{2}c_{3} \xi^2)}{c_{4}(\gamma_{1}c_{2}\eps x^{2}+\gamma_{1}^{2}\eps^{2}y^{2}+3c_{2}c_{3} \xi^2)^{2}}, 
\\
			\frac{j_y}{j_\Gamma}= -\frac{16\eps^{3} \xi^2\gamma_{1}^{3}c_{2}c_{3}xy}{c_{4}(\gamma_{1}c_{2}\eps x^{2}+\gamma_{1}^{2}\eps^{2}y^{2}+3c_{2}c_{3} \xi^2)^{2}}. 
		\end{gather}
	\end{subequations}
The supercurrent density at the instanton center is given by:
	\be\label{cursupression}
	j_x(\br=0)=j_{\Gamma}[1-\eta(T)\eps^2],
	\ee
with $\eta(T)=8\gamma_1^2/3c_2 c_4$.
		
Equation \eqref{cursupression} provides an estimate for the current $I_{\text{top}}(T)$ when $\mathbf{j}$ vanishes at the instanton center. 
This condition is believed to mark the second-order transition between topologically trivial (Boussinesq-like at $I>I_\text{top}$) and nontrivial (vortex-antivortex pair at $I<I_\text{top}$) instanton configurations \cite{SkvPol25}. Our numerics show that the coefficient $\eta(T)$ lies within a very narrow range $(15.93,16.00)$, corresponding to $I_{\text{top}}(T) \sim 0.94\,I_c(T)$. Hence, the estimate of the vortex-free region as approximately 6--7$\%$ of the critical current, obtained in Ref.~\cite{SkvPol25}, remains practically unchanged at arbitrary temperatures.

\begin{figure}
		\centering
		\includegraphics[width=\linewidth]{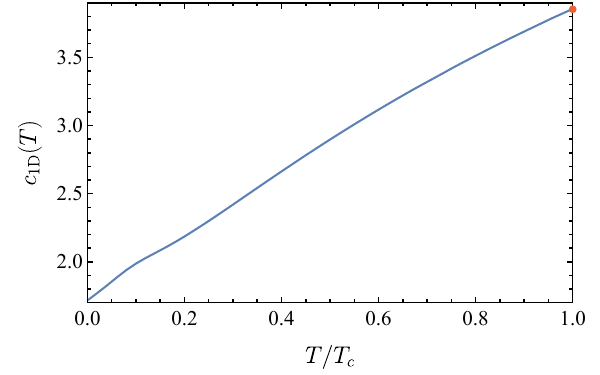}
		\caption{Temperature dependence of the coefficient $c_{1\text{D}}(T)$ in the activation barrier \eqref{ans1D} for a 1D wire. The red point corresponds to the LA analytical result in the GL region: $c_{1\text{D}}(T_c)=2^{25/4} 3^{-5/4}5^{-1} \approx 3.86$ \cite{LA}.}
		\label{fig:c1D}
\end{figure}

\subsection{Wires}

As a corollary of the developed formalism, one can easily obtain the phase-slip activation energy for 1D wires at $I\to I_c$ in the whole temperature range. In this case, the Boussinesq free energy \eqref{F-Bouss-dimless} lacks the term with the mixed derivative. Then after identifying $\bar x$ with time, it becomes formally equivalent to the action of a mechanical particle of unit mass with the coordinate $u=\bar f_{\bar x\bar x}$ in the potential $U(u)=-u^2/2-u^3/3$. Its bounce solution is $u(\bar x)=-(3/2)\sech^2 (\bar x/2)$ with $S=6/5$ per unit $\bar y$, and we arrive at
\be\label{wirebar}
  \Delta F_{1\text{D}}
  =
  \frac{16 c_1^{5/2} c_3^{1/2}}{15 c_4^2} \nu\Delta_{\G_c}^2 dw\xi ,
\ee
where $w$ is the wire width ($dw$ is its cross section).

Rewriting Eq.\ \eqref{wirebar} in terms of the GL condensation energy density \eqref{e-cond-GL}, we obtain for the phase slip barrier at $I\to I_c$:
	\be\label{ans1D}
	\Delta F_{1\text{D}}=c_{1\text{D}}(T) \eps_{\text{cond}}(T) dw \xi(T) (1-I/I_c)^{5/4} ,
	\ee
where
    \be
{c_{1\text{D}}(T)=\frac{28\zeta(3)}{15\pi^{2}}\frac{\gamma_1^{5/2}c_{3}^{1/2}}{c_{4}^{2}}\frac{\Delta_{\G_{c}}^{2}(T)}{(T_{c}-T)^{2}}} .
    \ee
The temperature dependence of $c_{1\text{D}}(T)$ is shown in Fig.~\ref{fig:c1D}. As $T\to T_c$, it agrees with the LA result \cite{LA} taken in the limit $I\to I_c$. 
In the mechanical analogy employed by LA, the saddle point solution is mapped onto a motion of $|\Delta|$ in an effective potential $U_\text{eff}(|\Delta|)$. Near the critical current, when the minimum of this potential is about to vanish, $U_\text{eff}$ becomes asymptotically equivalent (up to an appropriate rescaling and shift of variables) to our quadratic-cubic potential $U(u)$. Consequently, the bounce solution derived from $U(u)$ asymptotically coincides with that of Ref.\ \cite{LA}. At zero temperature, $c_\text{1D}(0)=1.72$.

    \begin{figure}[t]
		\centering
		\includegraphics[width=\linewidth]{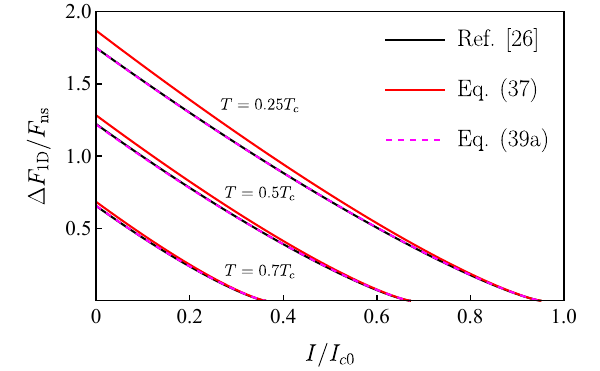}
\caption{Current dependence of the activation barrier $\Delta F_\text{1D}$ for a 1D wire at three values of $T/T_c$.
Shown are the numerical data of Ref.~\cite{SemDev} (black), our near-critical asymptotics \eqref{ans1D} (red), and the 
fit \eqref{F1D-Upsilon} (dashed). Currents are scaled by $I_{c0}$, the critical current at $T=0$.}
		\label{fig:semdevcomp}
	\end{figure}

Now we are in a position to compare our results, valid at $I\to I_c$, with  those of Semenov \emph{et al.}\ \cite{SemDev}, which were obtained for arbitrary currents by numerically solving the coupled Usadel equations \eqref{usadel} and the self‑consistency equation \eqref{selfcons}. 
Figure \ref{fig:semdevcomp} reproduces (black lines) the full current dependence of the activation barrier calculated in Ref.\ \cite{SemDev} for temperatures $T/T_c=0.25$, $0.5$, and $0.7$. 
The current is normalized to its zero-temperature critical value $I_{c0}\approx 0.47 \, \pi e \nu D^{1/2} \Delta_0^{3/2}(0) dw$ [where $\Delta_0(0)=1.76\,T_c$ is the zero-temperature gap in the absence of a current], and the free energy is normalized to $F_\text{ns}=\nu D^{1/2} \Delta_0^{3/2}(0) dw$.
Our expression \eqref{ans1D} is plotted as red curves for the same temperatures. It perfectly reproduces the asymptotic behavior of the numerical solution in the vicinity of $I_c$. This agreement provides mutual verification for both our results and those of Ref.~\cite{SemDev}.

Figure \ref{fig:semdevcomp} also demonstrates that Eq.\ \eqref{ans1D}---which is expected to be accurate only in the vicinity of $I_c$---is in fact a fairly good approximation for all currents down to zero. This feature is known in the GL region, where the exact current dependence of $\Delta F_\text{1D}$ can be fitted well (to within 2\% accuracy, even at $I=0$) by the leading asymptotic term near criticality, $(1-I/I_c)^{5/4}$ \cite{LA,Semenov2020}. Our comparison with Ref.~\cite{SemDev} indicates that the same behavior extends to arbitrary temperatures $T<T_c$.

Motivated by the small discrepancy between $\Delta F_\text{1D}$ and its asymptotic expansion \eqref{ans1D}, we can propose a practical formula that approximates the activation barrier for all $I$ and $T$:
\begin{subequations}
\label{F1D-Upsilon}
\be
  \Delta F_\text{1D}
  \approx
  \Upsilon(I,T)
  c_{1\text{D}}(T) \eps_{\text{cond}}(T) dw \xi(T) (1-I/I_c)^{5/4} ,
\ee
with the correction function
\be
\label{Upsilon}
  \Upsilon(I,T)
  =
  1-h(T/T_c) (1-I/I_c)^{1/2} (1+I/I_c)^{7/4-4\tau/3} ,
\ee
\end{subequations}
where $h(T/T_c)=0.022+0.056 \, \tau$ and $\tau = 1-T/T_c$. The last factor in Eq.\ \eqref{Upsilon} accounts for the temperature dependence of the curvature of $\Delta F_\text{1D}$. 
The approximation \eqref{F1D-Upsilon}, shown as dashed lines in Fig.\ \ref{fig:semdevcomp}, is nearly indistinguishable from the numerical data of Ref.\ \cite{SemDev}.

\subsection{Comment on the current-source work}

Finally, we address the issue of the work performed by the current source during the evolution from the uniform supercurrent state to the saddle-point configuration. 
Since this process is considered at a constant current bias $I$, the appropriate thermodynamic potential includes an additional contribution \cite{McCumber}:
\be\label{dFfull}
  \Delta F = F_{\text{sad}} - F_{\text{uni}} - \frac{I \Delta \phi}{2e} ,
\ee
where $\Delta \phi = \phi(x=+\infty) - \phi(x=-\infty)$ is an extra phase difference between the edges of the sample accumulated in the saddle-point state [see Eq.\ \eqref{deltaphi}].
In the 1D geometry, $\Delta\phi$ is finite \cite{LA,McCumber}, and the current-source work should be taken into account. However, the expansion of $F_{\text{sad}} - F_{\text{uni}}$ itself starts with the term linear in the phase gradient $\phi_x\sim \eps$:
\be
\label{F1}
  F^{(1)} = \pi\nu DTd \k\sum_{n}\sin^{2}\t_{n}\int dx \, dy \, \phi_{x} ,
\ee
which, at the uniform solution, can be transformed into the counterterm $F^{(1)} = I \Delta \phi/2e$, completely compensating the work of the current source. This property has allowed us to find the phase-slip barrier as the saddle-point value of the functional \eqref{F23}, which contains only the terms scaling as $\eps^2$ and $\eps^3$.

The same argument applies to the 2D saddle point obtained in Ref.\ \cite{SkvPol25}. 
In this case, Eq.\ \eqref{dFfull} generalizes to
\be
\label{dFfull-gen}
  \Delta F
  =
  F_{\text{sad}} - F_{\text{uni}} 
  -
  \frac{\bj_0}{2e} \int d\br \, \nabla \phi ,
\ee
where $\bj_0$ is the uniform current density at infinity.
Although in 2D the extra phase vanishes at infinity as $\phi(\br)\propto \bar x/(\bar x^2+\bar y^2)$, its $1/r$ tail nonetheless yields a finite contribution from the current source, like in the 1D case.
However the linear-in-$\eps$ term \eqref{F1} (suitably generalized to include $y$ dependence) cancels the resulting current-source work, thereby justifying the retention of terms scaling as $\eps^2$ and $\eps^3$.

\section{Summary and conclusion}
\label{S:Summary}

In this work we have carried out a mostly analytical study of thermal phase slips in thin superconducting films near the critical current at arbitrary temperatures. In the vicinity of $I_c$, the saddle-point configuration is topologically trivial, with a nonvanishing order parameter. In the GL region near $T_c$, the instanton profile together with the scaling of its size was obtained by one of the authors~\cite{SkvPol25}. We generalize the main result of Ref.\ \cite{SkvPol25} to arbitrary $T<T_c$, working within the Usadel description of dirty superconductors. Using the scaling of the instanton parameters obtained in the GL region, we perform a gradient expansion in the Usadel equations and, similarly to Ref.\ \cite{SkvPol25}, derive the leading-order free-energy functional in an exactly solvable Boussinesq form. 

Our approach is built upon two core ingredients. First, using the divergence of both sizes ($L_x$ and $L_y$) of the optimal fluctuation as $I\to I_c$, we perform the gradient expansion of the Usadel equations. This procedure results in the free energy \eqref{F23} expressed in terms of the deviations of the order parameter (its modulus $\delta$ and phase $\phi$) from the uniform-current solution.
Second, we use the current conservation equation to reduce these two fields, $\delta(\br)$ and $\phi(\br)$, to a single real field $f(\br)$ [see Eq.~\eqref{reduction}]. This step directly transforms the free energy into the Boussinesq form \eqref{Bouss1} written in terms of $f(\br)$. The possibility of performing such a reduction is related to the effective one-dimensionality of the free-energy functional as $I\to I_c$, which in turn is connected to the strong anisotropy of the instanton, with $L_y\gg L_x$.

Although our introduction of $f(\br)$ does not rely on the stream function used in Ref.~\cite{SkvPol25}, both methods share current conservation as their physical basis. The resulting functions coincide only to the required leading orders, which is sufficient for deriving the free energy in the Boussinesq form.

Our main result comprises analytical expressions for the coefficients $c_{2\text{D}}(T)$ and $c_{1\text{D}}(T)$, which govern the asymptotics \eqref{ans2D} and \eqref{ans1D} of the activation barrier near the critical current in 2D and 1D, respectively. The coefficients $c_{2\text{D}}(T)$ and $c_{1\text{D}}(T)$ are expressed as Matsubara sums over the spatially uniform spectral angles that solve the Usadel equation in the presence of a uniform supercurrent.
Evaluating these sums numerically is considerably less demanding than determining the nonuniform instanton configuration from the coupled Usadel and self-consistency equations.
Figures \ref{fig:c2D} and \ref{fig:c1D} demonstrate that both $c_\text{2D}(T)$ and $c_\text{1D}(T)$ decrease with temperature in a nearly linear fashion, with $c_{\text{D}}(0)\approx 0.5 \, c_{\text{D}}(T_c)$.

For a 1D wire, the activation barrier near criticality, $\Delta F_\text{1D}(T,I\to I_c)=E_\text{1D}(T)(1-I/I_c)^{5/4}$, is consistent with the numerics of Semenov \emph{et al.}\ \cite{SemDev}, mutually validating both calculations. Similar to the LA result in the GL region, this near-$I_c$ form remains a good approximation to the full barrier $\Delta F_\text{1D}(T,I)$ for all temperatures and currents. Combining our results with those of Ref.\ \cite{SemDev}, we propose the interpolation formula \eqref{F1D-Upsilon}, which reproduces the barrier across the entire range $T<T_c$ and $I<I_c(T)$ to within a 1\% accuracy.

In the 2D case, the Boussinesq approximation is accurate only in a narrow region near the critical current. At $I=I_\text{top}(T)$, the order parameter vanishes at the instanton center, signaling a second-order topological transition between vortex-free ($I>I_\text{top}$) and vortex-antivortex ($I<I_\text{top}$) saddle-point configurations \cite{Vodolazov,SkvPol25}. Notably, the estimate $I_\text{top}(T)\approx0.9 I_c(T)$ obtained in Ref.\ \cite{SkvPol25} is nearly temperature independent.

	\acknowledgments

We are grateful to A. V. Polkin and G. E. Volovik for stimulating discussions and to A. V. Semenov for sharing his numerical data.

	\appendix

\section{Coefficients in the free energy \eqref{F23}}
\label{S:Appendix A}

The free energy expansion \eqref{F23} in terms of the fields $\delta$ and $\phi$ is characterized by the set of coefficients $a_i$ and $b_i$, which are obtained by substituting Eqs.\ \eqref{tx-via-df} into Eq.~\eqref{freeenergy}. The resulting expressions have the following form: 
\begin{gather}
\label{a1}
  a_1
  =
  \frac{\pi T}{\D}
  \sum_n\frac{(\D - \G \sinn)\sinn^2}{\mathcal{E}_n} ,
\\
\label{a2}
  a_2
  =
  \frac{2\pi D \kappa T}{\xi\D}
  \sum_n\frac{\sinn\cosn^2}{\mathcal{E}_n} ,
\\
\label{a3}
  a_3
  =
  \frac{\pi D T}{2\xi^2\D^{2}}
  \sum_n\sinn^2
  \biggl[
    1-\frac{4\G\cosn^2}{\mathcal{E}_n}
  \biggr] ,
\\
\label{b1}
  b_1
  =
  \frac{\pi D T}{2\xi^2\D^{2}} \sum_n\sinn^2 ,
\\
\label{b2}
  b_2
  =
  \frac{\pi D T}{\xi^2\D}
  \sum_n \sinn^2\cosn^2
  \biggl[
    \frac{1}{\mathcal{E}_n}
  - \frac{2\G (5\cosn^2-3)}{\mathcal{E}_n^2} 
  - \frac{6\G^{2}\sinn^2\cosn^2}{\mathcal{E}_n^3}
  \biggr] ,
\\
\label{b3}
  b_3
  =
  -
  \frac{\pi D^2 \kappa T}{\xi^3 \D^{2}}
  \sum_n \sinn^2\cosn^2
  \biggl[
      \frac{1}{\mathcal{E}_n}
    - \frac{2\G (2\cosn^2-1)}{\mathcal{E}_n^2}
    - \frac{2\G^{2}\sinn^2\cosn^2}{\mathcal{E}_n^3}
  \biggr] ,
\\
\label{b4}
  b_4
  =
  \pi\D T \sum_n
  \sinn\cosn^2
  \biggl[ 
    \frac{1}{\mathcal{E}_n^2} - \frac{\G\cosn^2}{\mathcal{E}_n^2}
  \biggr] ,
\\
\label{b5}
  b_5
  =
  \frac{\pi D \kappa T}{\xi}
  \sum_n \cosn^2
  \biggl[ 
    \frac{4\cosn^2-3}{\mathcal{E}_n^2}
  + \frac{3 \G \sinn^2 \cosn^2}{\mathcal{E}_n^3}
  \biggr] ,
\\
\label{b6}
  b_6
  =
  \frac{\pi D T}{2\xi^2\D} \sum_n
  \cosn^2 \frac{\D-4\G \sinn \cosn^2}{\mathcal{E}_n^2} ,
\\
\label{b7}
  b_7
  =
  -
  \frac{\pi D^2 T}{4\xi^4\D^3}
  \sum_n
  \sinn^2
  \biggl[
    \sinn - \frac{8 \G\sinn\cosn^2}{\mathcal{E}_n} 
    - \frac{4\G\cosn^2 (\D- 4 \G \sinn \cosn^2)}{\mathcal{E}_n^2}
  \biggr] ,
\\
\label{b8}
  b_8
  =
  \frac{\pi D^2 \kappa T}{\xi^3\D^{2}}
  \sum_n
  \sinn\cosn^2
  \biggl[
    \frac{\sinn}{\mathcal{E}_n} 
  + \frac{\D- 4 \G \sinn \cosn^2}{\mathcal{E}_n^2}
  \biggr] .
\end{gather}
Here $\sinn \equiv \sin\theta_n$ and $\cosn \equiv \cos\theta_n$, where the spectral angle $\theta_n \equiv \t_{\G n}$ satisfies the Usadel equation \eqref{uniformusadel} with $\Gamma=D\kappa^2/2$ and $\Delta \equiv \D_\G$ obtained self-consistently for a given temperature. Denominators in Eqs.\ \eqref{a1}--\eqref{b8} contain (half of) the cooperon mass,
\be
  \mathcal{E}_n
  \equiv
  \epsilon_n \cosn + \Delta\, \sinn + \Gamma (\cosn^2-\sinn^2)
  =
  \frac{\D - \G \sinn^3}{\sinn},
\ee
where the last equality, which follows from Eq.\ \eqref{uniformusadel}, may be used to analytically simplify intermediate expressions.

According to Sec.\ \ref{S:Expansion}, the coefficients $c_2$, $c_3$ and $c_4$ in the Boussinesq free energy \eqref{Bouss1} are expressed in terms of $a_i$ and $b_i$ evaluated exactly at criticality. On the other hand, to determine the coefficient $c_1$, which vanishes as $\gamma_1\sqrt{1-I/I_c}$, one needs to know how $a_i$ depend on $I$ in the vicinity of $I_c$.

\emph{In the GL region} ($T\to T_c$), the order parameter is small ($\Delta \propto \tau^{1/2}$), and the spectral angle can be expanded in $\Delta$. In the leading order, $\theta_n \approx \Delta/\epsilon_n$, 
and $\Gamma\propto\Delta^2\ll\Delta$ can be neglected under the sums of Eqs.\ \eqref{a1}--\eqref{b8}. As a result, one can easily calculate these expressions explicitly in the leading order $O(\tau)$:
\begin{subequations}\label{aibiGL}
\begin{align}
\label{ai-GL}
  \begin{pmatrix} 
    a_1 & a_2 & a_3
  \end{pmatrix}
  & = 
  \begin{pmatrix} 
    2 (1-A^2) & 4A & 1
  \end{pmatrix}
  \tau ,
\\
\label{bi-GL}
  \begin{pmatrix} 
    b_1 & b_2 & b_3 & b_4 \\ 
    b_5 & b_6 & b_7 & b_8 
  \end{pmatrix}
  & =
  \begin{pmatrix} 
    1 & 2 & 0 & 2(1-A^2) \\ 
    2A & 1 & 0 & 0 
  \end{pmatrix}
  \tau ,
\end{align}
\end{subequations}
where $A=\kappa\xi$.

Note that the coefficients $b_3$, $b_7$ and $b_8$ must vanish at the GL level because they multiply terms in the free energy expansion \eqref{F23} that contain more than two derivatives. Such terms do not appear in the GL expansion, and the corresponding coefficients---which are indeed zero in Eq.\ \eqref{bi-GL}---are subleading, scaling as $\tau^2$. 

Note also the relation $4 a_1 a_3 - a_2^2 = 8 \tau^2 (1-3A^2) \propto 1-I/I_c$, which follows from Eq.\ \eqref{ai-GL}. This illustrates, in the GL region, the general mechanism for $c_1$ vanishing at the critical current (see Appendix \ref{S:Appendix B}).

\section{Relation between $a_i$ at $I_c$}
\label{S:Appendix B}
	
Here we prove the identity $4a_1 a_3 - a_2^2 = 0$ at the critical current. According to Eq.\ \eqref{c1}, this property is responsible for the scaling of the coefficient $c_1\propto\sqrt{1-I/I_c}$ in the Boussinesq free energy \eqref{Bouss1}.

The critical current is obtained as a maximum of $j(\Gamma)$. 
With the help of Eqs.\ \eqref{current} and \eqref{Gamma-def}, we rewrite the condition $\partial j/\partial \G=0$ in the form
	\be\label{djdg}
	\sum_n\sin^2{\t_n} =
	-4\G \sum_n\sin\t_n\cos\t_n \frac{\partial \t_n}{\partial\G} .
	\ee
The derivative $\partial\t_n/\partial\G$ should be determined simultaneously with $\partial\Delta/\partial\G$ from the system of the Usadel equation \eqref{uniformusadel} and self-consistency equation \eqref{selfcons}. After some algebra, we obtain
\begin{subequations}
\begin{gather}
\label{dertheta}
  \frac{\partial \theta_n}{\partial \G}
  =
  - \frac{\sin \t_n \cos\t_n}{\mathcal{E}_n} 
  + \frac{\cos\t_n}{\mathcal{E}_n} 
  \frac{\partial\Delta}{\partial\Gamma} ,
\\{}
  \frac{\partial\Delta}{\partial\Gamma}
  =
  - \frac{\sum_n \sin \t_n \cos^2\t_n/\mathcal{E}_n}
  {\sum_n (\Delta - \G \sin \t_n) \sin^2 \t_n/\mathcal{E}_n\D}
  .
\end{gather}
\end{subequations}
Substituting Eq.\ \eqref{dertheta} into Eq.\ \eqref{djdg} provides an explicit expression for $\sum_n \sin^2 \theta_n$ that holds exactly at the critical current. Applying it to transform Eq.\ \eqref{a3} for $a_3$ then yields the desired identity:
\be
  a_3(I_c)=\frac{a_2(I_c)^2}{4a_1(I_c)} ,
\ee
which is valid for all temperatures.

\bibliography{references}

\end{document}